# Jahn-Teller distortion induced charge-ordering in CE phase of manganites

S. Dong, S. Dai, X.Y. Yao, K.F. Wang, C. Zhu and J. –M. Liu*

*Laboratory of Solid State Microstructures, Nanjing University, Nanjing 210093, China*

*International Center for Materials physics, Chinese Academy of Sciences, Shenyang, China*

**[Abstract]** The charge order of CE phase in half-doped manganites is studied, based on an argument that the charge-ordering is caused by the Jahn-Teller distortions of $MnO_6$ octahedra rather than Coulomb repulsion between electrons. The quantitative calculation on the ferromagnetic zigzag chain as the basic structure unit of CE phase within the framework of two-orbital double exchange model including Jahn-Teller effect is performed, and it is shown that the charge-disproportionation of Mn cations in the charge-ordered CE phase is less than 13%. In addition, we predict the negative charge-disproportionation once the Jahn-Teller effect is weak enough.

---

\* Corresponding author. E-mail: liujm@nju.edu.cn

Manganites as typical strongly correlated electron systems have been extensively studied over the last decade because of the observed colossal magnetoresistance effect (CMR). The appearance of various phases in manganites upon the change of doping, temperature and applied external field is repeatedly confirmed, and the abundant physical phenomena associated with phase separation (PS) of manganites reserve special motivations for fundamental and applied research activities.[1] Here, we focus on the microscopic mechanism for the fairly complex CE phase in half-doped manganites (with the general chemical formula $R_{1/2}A_{1/2}MnO_3$, where R and A are rare- and alkaline-earth cations, respectively). This topic is still under discussion although quite a lot of effort has been made so far.[2,3] In the conventional viewpoint, charge-ordering (CO) in manganites means an ordered arrangement of $Mn^{3+}/Mn^{4+}$ cations in some special doping density, e.g. divalent doping $x$=1/2, 1/3, and 1/4. The CE phase at $x$=0.5 has equal amount of $Mn^{3+}/Mn^{4+}$ aligned in a checkerboard pattern in $x$-$y$ plane (Fig.1(a)), with charges stacked along $z$-axis (Fig.1(b)).[4,5] The prominent microscopic character of the CE phase is the appearance of CO[6] and ferromagnetic (FM) zigzag stripes (Fig.1(a)),[7] which is believed to be the effect of the competition among various interactions.

Theoretical[8] and experimental[9] evidences available so far allow us to argue that the above scenario is oversimplified, and our conventional understanding of the CE phase in simple mixed-valent oxides, e.g. $Fe_3O_4$ and $Ti_2O_3$, is challenged.[10] A scenario where Mn cation and even O anion have non-integer valences is suggested.[10] Quantitatively, the charge-disproportionation factor $\delta$, defined as $n_2$-$n_1$ where $n_2/n_1$ is the maximum/minimum valence of mixed-valent cation, can be used as a criterion measure of the CO state. For example, for the CE type manganites $\delta$=1 is conventionally believed, because $Mn^{3+}$ is richer than $Mn^{4+}$ for one $e_g$ electron, similar to the case of $Fe^{2+}/Fe^{3+}$ in $Fe_3O_4$,[11] while for metallic manganites such as $La_{0.7}Sr_{0.3}MnO_3$, where $e_g$ electrons are communized and no CO can exist stably, $\delta$=0 is understood because all Mn cations are same in charge.

Recently, Brink $et$ $al$ studied in detail the charge- and orbital-ordering of CE phase[12] using a double exchange (DE) model on a FM zigzag chain (Fig.1(c)), since the FM zigzag chain is the basic periodic structure unit of CE phase. The detailed calculation can also be found in Ref.[1]. It was found that the orbital occupancy at bridge sites is $|3x^2 - r^2 > / |3y^2 - r^2 >$ and that at corner sites is $\frac{\sqrt{3}}{2}|x^2 - y^2 > \pm \frac{1}{2}|3z^2 - r^2 >$, as shown in Fig.1(d). Therefore, the CE phase is orbital-ordered, and charges are homogeneously distributed (each site no matter the corner or bridge one has 1/2 electron and thus $\delta$=0). However, this model predicts significant charge-disproportionation when Coulomb interaction $U$ between electrons on different orbitals of the same site is taken into account. On the bridge site, one orbital is always empty, and the Coulomb repulsion is ineffective. On the corner site, however, both orbitals are partially occupied, so that the charge is pushed away from the correlated corner sites to the uncorrelated bridge sites, causing the CO state. In this model, factor $\delta$ increases monotonously with $U$ and approaches its maximum limit 18.5% as $U \rightarrow \infty$. A subsequent investigation[13] studied the charge- and orbital-ordering in CE phase, considering more complex interactions such as Coulomb interaction between electrons on nearest-neighbor (NN) sites and Jahn-Teller (JT) distortions.

However, the above claimed essential role of the Coulomb repulsion in causing CO remains specious. The Hamiltonian used in Ref.[12], $H_C = U \sum_{\alpha < \beta, i} n_{i,\alpha} n_{i,\beta}$, may not describe the Coulomb interaction in an appropriate manner, because the operator $n_{i,\alpha} n_{i,\beta}$ depends on the choice of the basic vectors (orbitals $\alpha$ and $\beta$). Here, the invariance associated with a linear transform of basic vectors is broken, which becomes a paradox if one notices that the linear transform of basic vectors would not affect the physical result. This paradox is due to the artificial division of 1/2 electron in a site into two orbitals. Consider a vector $\boldsymbol{r} = x\boldsymbol{i} + y\boldsymbol{j}$ where basic vectors $\boldsymbol{i}$ and $\boldsymbol{j}$ are perpendicular to each other. The product $xy$ is dependent on the choice of $\boldsymbol{i}$ and $\boldsymbol{j}$. The invariant under the linear transform is module $|\boldsymbol{r}|$ whose physical sense is irrelevant to the transform operation. In addition, if the Coulomb interaction between $3/8\alpha$ electron and $1/8\beta$ electron in a corner site is $H_C$, one has no reason to omit the interaction between $1/4\alpha$ electron and another $1/4\alpha$ electron in a bridge site. It is not easy to compare the Coulomb energy of 1/2 electron in $\frac{\sqrt{3}}{2}|x^2 - y^2 > \pm \frac{1}{2}|3z^2 - r^2 >$ forms and in $|3x^2 - r^2 > / |3y^2 - r^2 >$ forms without detail calculation. Considering the

self-energy of electron, one has $H_C = U' \sum_i n_i^2$ where $U'$ is the Coulomb energy factor, if the shape factor of electron cloud can be neglected. This new Hamiltonian is invariable under linear transform of the basis vectors, and the charge distribution should be homogeneous (with equal $n_i$ at all sites) in order to lower the electrostatic energy. For nanoscale phase separation in manganites, the Coulomb interaction has the similar effect so that the sizes of PS clusters with different charge densities are limited in nanometer scale.[1] Consequently, the idea that the Coulomb interaction induces charge-disproportionation in CE phase is misleading somehow.

Here we propose a mechanism for the charge-ordering in CE phase: JT distortions as the main origin to cause the CO. In the common knowledge, the coupling of JT phonons with $e_g$ electrons is significant in the physics of manganites.[14] For metallic manganites, the lattice distortions are absent and $e_g$ electrons are delocalized. Consequently, the charge density is homogeneous. However, for the strong JT distortion case, charge-ordered states are often stable,[15] which evidences the strong dependence of CO states on the JT distortion. The direct consequence of the JT distortions is to split the two-fold degenerate $e_g$ energy levels into two orbitals: $a$ (higher energy) and $b$ (lower energy), as shown in Fig.2(a). The energy difference between $a$ and $b$ is $E_{JT}$.

We employ the DE model with JT effect to study the CO behaviors in CE phase. The detail of the calculation is similar to earlier ones.[1,12] A FM zigzag chain with four sequential sites is considered as a periodic structure unit of CE phase (Fig.1(c)), with Bloch phase factor $e^{ikr}$ added to the wave functions of $e_g$ electrons.[16] The basic vectors are the original wave functions of orbitals $a$ and $b$ on the four sites. Simplifying the JT distortions to a static effect of $e_g$ levels split, one has the Hamiltonian of $e_g$ electron:

$$H = -\sum_{<i,j>,\alpha,\beta} t_{\alpha\beta}^d c_{i\alpha}^+ c_{j\beta} + E_{JT} \sum_i (n_{ia} - n_{ib})  \qquad (1)$$

where $c_{i\alpha}^+ / c_{j\beta}$ creates/annihilates an electron on site $i/j$ in orbital $\alpha/\beta$ and $<i,j>$ denotes a NN pair; $n$ is the number operator; $t_{\alpha\beta}^d$ is the hopping integral of orbitals $\beta$-to-$\alpha$ along direction $d$ (Fig.2(b)). The hopping of $e_g$ electron is constrained along the FM zigzag chain in $x$-$y$ plane. It is known the values of $t^d_{\alpha\beta}$ are:[1]

$$t^x = \frac{t_0}{4}\begin{pmatrix} 3 & -\sqrt{3} \\ -\sqrt{3} & 1 \end{pmatrix}, \quad t^y = \frac{t_0}{4}\begin{pmatrix} 3 & \sqrt{3} \\ \sqrt{3} & 1 \end{pmatrix} \tag{2}$$

We use $t_0$ as the energy unit, so the total Hamiltonian and wave functions can be expressed as:

$$\frac{1}{4}\begin{pmatrix} 2E_{JT} & 0 & -3e^{ik} & \sqrt{3}e^{ik} & 0 & 0 & -3e^{-ik} & -\sqrt{3}e^{-ik} \\ 0 & -2E_{JT} & \sqrt{3}e^{ik} & -e^{ik} & 0 & 0 & -\sqrt{3}e^{-ik} & -e^{-ik} \\ -3e^{-ik} & \sqrt{3}e^{-ik} & 2E_{JT} & 0 & -3e^{ik} & \sqrt{3}e^{ik} & 0 & 0 \\ \sqrt{3}e^{-ik} & -e^{-ik} & 0 & -2E_{JT} & \sqrt{3}e^{ik} & -e^{ik} & 0 & 0 \\ 0 & 0 & -3e^{-ik} & \sqrt{3}e^{-ik} & 2E_{JT} & 0 & -3e^{ik} & -\sqrt{3}e^{ik} \\ 0 & 0 & \sqrt{3}e^{-ik} & -e^{-ik} & 0 & -2E_{JT} & -\sqrt{3}e^{ik} & -e^{ik} \\ -3e^{ik} & -\sqrt{3}e^{ik} & 0 & 0 & -3e^{-ik} & -\sqrt{3}e^{-ik} & 2E_{JT} & 0 \\ -\sqrt{3}e^{ik} & -e^{ik} & 0 & 0 & -\sqrt{3}e^{-ik} & -e^{-ik} & 0 & -2E_{JT} \end{pmatrix}\begin{pmatrix} c^+_{1a} \\ c^+_{1b} \\ c^+_{2a} \\ c^+_{2b} \\ c^+_{3a} \\ c^+_{3b} \\ c^+_{4a} \\ c^+_{4b} \end{pmatrix} = E\begin{pmatrix} c^+_{1a} \\ c^+_{1b} \\ c^+_{2a} \\ c^+_{2b} \\ c^+_{3a} \\ c^+_{3b} \\ c^+_{4a} \\ c^+_{4b} \end{pmatrix} \tag{3}$$

We diagonalize this Hamiltonian exactly and the eigen energy is shown in Fig.3. It is shown that the energy bands in the reduced zone of the $e_g$ electrons in a FM zigzag chain of CE phase are constrained. The results at $E_{JT}$=0 are the same as the ones reported earlier.[1,12] However, for $E_{JT}$>0, e.g. $E_{JT}$=3.8$t_0$, the energy bands show quite different pattern referring to the case of $E_{JT}$=0. The two low and fully filled bands become lower, and the two high empty bands become higher. The middle band, which is four-fold degenerate at $E_{JT}$=0, splits into four branches.

Besides the energy band, we also calculate the charge distribution of the FM zigzag chain. In the ground state, $e_g$ electrons fully occupy the two lowest energy bands. The charge occupancy of orbital $a$/$b$ in corner/bridge site can be obtained by integrating the eigen wave functions of the two occupied bands. Fig.4(a) shows the charge occupancy as a function of $E_{JT}$ in each orbital/site. From Fig.4(a), we can evaluate (1) the average probabilities $n_{oa}$ (so as $n_{ob}$) over all sites to find an electron in each orbital $a$ (so as $b$), as shown in Fig.4(b); (2) the charge distribution $n_b$ (or $n_c$) in bridge (or corner) site, as shown in Fig.4(c). Then the charge-disproportionation $\delta=n_b-n_c$ as a function of $E_{JT}$ is depicted in Fig.4(d).

We partition the $E_{JT}$-dependence of the charge occupancy and disproportionation shown in Fig.4 into three regions. The first region refers to the small $E_{JT}$ case, i.e. $E_{JT}/t_0 \in [0,1]$. As $E_{JT}=0$, $e_g$ electrons prefer orbital $a$ rather than orbital $b$ ($n_{oa}>n_{ob}$) because of DE interaction ($t_{aa}=3t_{bb}$ both in $x$ and $y$ directions), and charge is homogenous at corner and bridge sites ($\delta=0$). The occupancy of orbital $a$ decreases with increasing $E_{JT}$ from zero, leading to $n_{oa}<n_{ob}$ at a threshold $E_{JT}/t_0 \approx 0.65$. What is interesting here is that a negative $\delta$ appears within $E_{JT}/t_0 \in [0,1]$ due to the even larger charge density at corner site than that at bridge site ($n_c>n_b$), although the absolute value of $\delta$ is no more than 2%. In the second region where $E_{JT}/t_0 \in [1, 3.8]$, $n_b$ is always larger than $n_c$ and $\delta$ increases monotonously with $E_{JT}$. It is identified that $\delta$ reaches its maximum value 12.7% at $E_{JT}=3.8t_0$. For a reference, experiment of X-ray resonant scattering on $Nd_{0.5}Sr_{0.5}MnO_3$ single crystal gave a value of $\delta=16\%$.[8] Quite good consistency between the calculation and experiment is shown, and the small difference between them can be smeared by considering more delicate interactions appeared in real material systems. Due to the large JT split of $e_g$ energy, $e_g$ electrons prefer the lower energy orbitals $b$ rather than the higher energy orbitals $a$, and the value of ($n_{ob}-n_{oa}$) also increases monotonously with $E_{JT}$. In the last region where $E_{JT}/t_0>3.8$, orbital $a$ becomes almost empty with a half-filled orbital $b$. The DE is seriously suppressed since there is only approximative one orbital degree of freedom, so the factor $\delta$ decreases with $E_{JT}$ slowly. In highlighting the above discussion, the JT distortion changes significantly the orbital occupancy: suppression of orbital $a$ and preference of orbital $b$. Due to the difference of orbital distribution between the corner and bridge sites, the influence of the JT distortion on the corner and bridge sites is quantitatively different, which causes the charge-disproportionation, or so called the CO state. Here the factor $\delta$ is far smaller than the conventionally expected value, so the present study provides a revised CO scenario.

However, what should be noted is that the FM zigzag CE phase is stable only in a narrow region of the phase diagram of manganites. Furthermore, the DE energy unit $t_0$ is also correlated with the JT distortion, because the DE process strongly depends on the angles of Mn-O-Mn bonds which lie on the mismatch of ions radii. The DE is restrained in the strong distortion case (large $E_{JT}$) and corresponding $t_0$ is smaller than the normal situation. So here $E_{JT}\sim3\text{-}4t_0$ is still acceptable, referring to the large $U/t_0$ taken in Ref.[12] which will destroy the stability of CE phase, as commented by Shen.[17]

In summary, the effect of Jahn-Teller distortion on the charge-ordering in the CE phase of manganites has been investigated. It has been revealed that the Coulomb interaction prefers a homogeneous charge distribution rather than charge-ordering state. The detailed calculation on energy bands and eigenstates of $e_g$ electrons has demonstrated that the Jahn-Teller distortion is responsible for the charge-ordering behaviors, with no more than 13% charge-disproportionation. Our result supports the current doubt on mixed-valent oxides, at least on half-doped manganites.

**Acknowledgement**

This work was supported by the Natural Science Foundation of China (50332020, 10021001) and National Key Projects for Basic Research of China (2002CB613303, 2004CB619004).

**Figure Caption:**

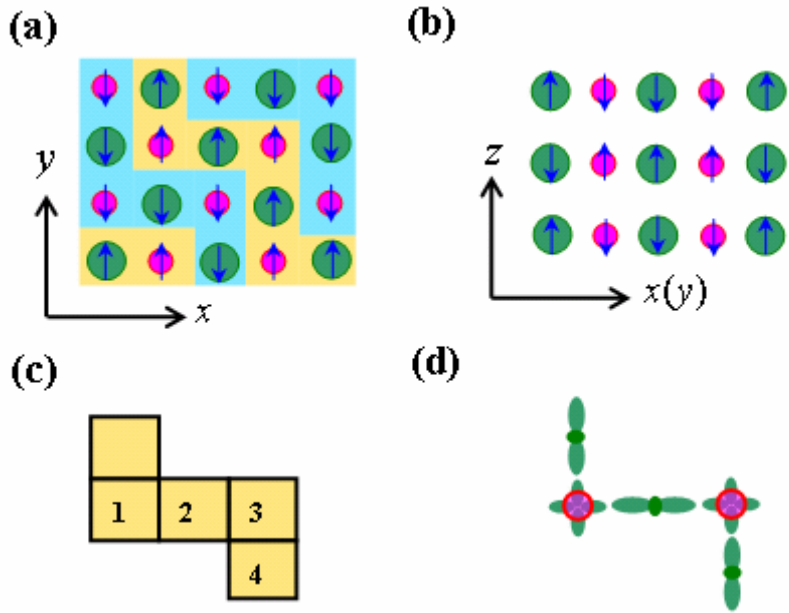

**FIG.1.** So called CE phase. (a) Checkerboard arrangement of Mn cations in $x$-$y$ plain (circles with different radii represent different valences and arrows in circles represent spins of $3d$ electrons). The ferromagnetic zigzag chain is shown as the colored stripes. (b) Charges stacked arrangement of Mn cations in $x(y)$-$z$ section, with antiferromagnetic coupling along $z$-axis between nearest neighbour $x$-$y$ layers. (c) A periodic unit of the CE phase: 4 sequential sites in a FM zigzag chain. It is cut from the one colored stripe in (a). The sites 1, 3 are called corner sites, and 2, 4 are bridge sites. (d) The orbital order of the FM zigzag chain.

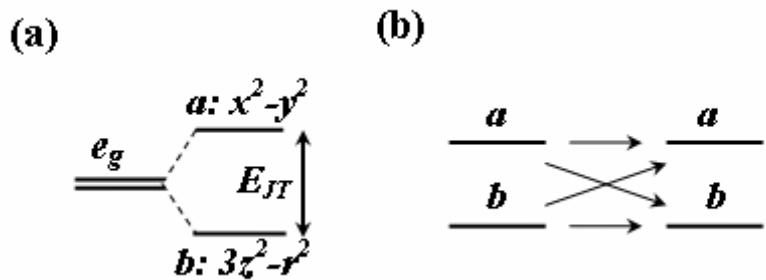

**FIG. 2.** (a) With Jahn-Teller distortions of $MnO_6$ octahedra, the two-fold degenerate $e_g$ energy level splits into two levels $a$ ($|x^2 - y^2>$) and $b$ ($|3z^2 - r^2>$) with the energy difference $E_{JT}$. (d) Two-orbital double exchange process.

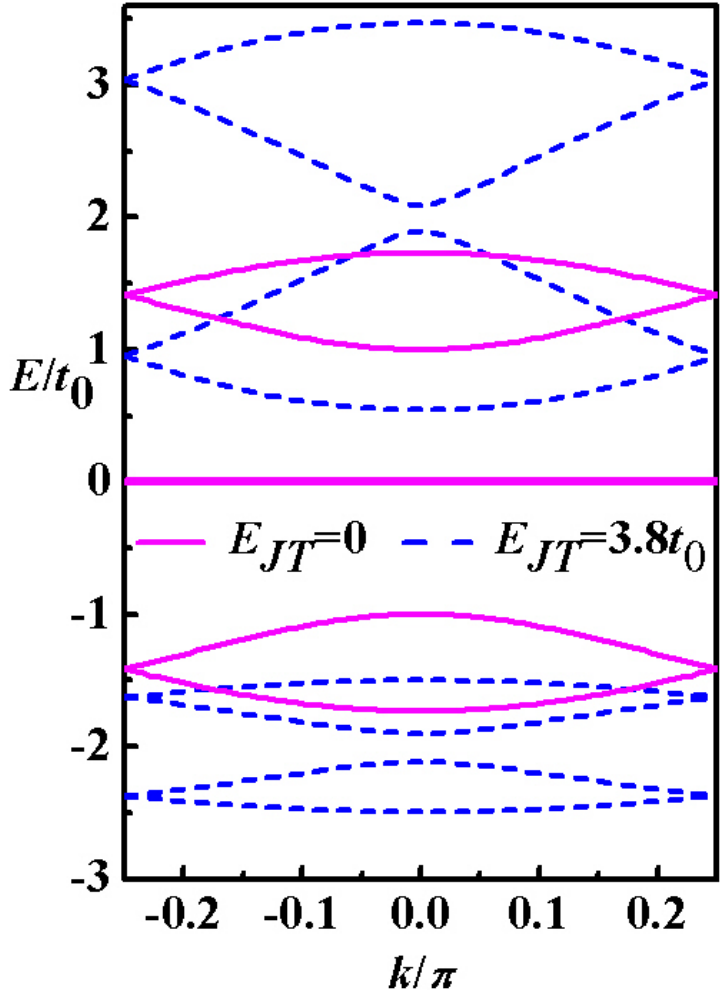

**FIG. 3.** Energy band structure for the zigzag chain in the reduced momentum zone [–π/4, π/4]. The solid curves are for the $E_{JT}$=0 case, while the broken curves represent $E_{JT}$=3.8$t_0$ case. The $E_{JT}$=0 case was studied by Brink *et al*[12] first, in which the flat $E$=0 band is four-fold degenerate. There are only the lower two bands of total eight are fully filled in ground state, while other six bands are empty. So the system is a band insulator due to the energy gap between filled bands and empty bands.

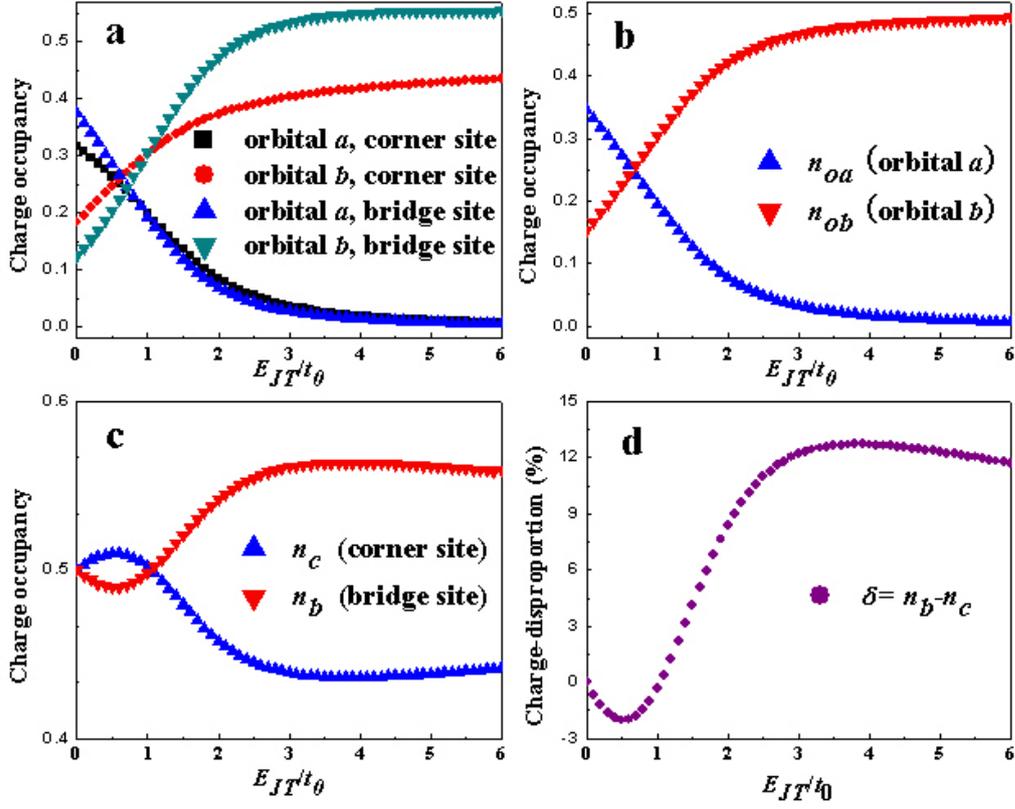

**FIG. 4.** Effects of Jahn-Teller split energy $E_{JT}$. (a) Charge occupancy of each orbital/site. (b) Charge occupancy of orbitals $a$ and $b$ in all sites (average value of the corner and bridge sites). (c) Charge occupancy of corner and bridge sites. (d) Charge-disproportionation $\delta$. The maximum $\delta$ 12.7% is obtained at $E_{JT}$=3.8$t_0$.